\documentclass[journal]{IEEEtran}
\usepackage{cite}

   \usepackage[dvips]{graphicx}
\usepackage[cmex10]{amsmath}
\usepackage{array}

\usepackage[tight,footnotesize]{subfigure}
\usepackage{url}
\usepackage{amsthm}

\begin{document}
%
\title{A Generalised Directional Laplacian Distribution: Estimation, Mixture Models and Audio Source Separation}
%
%
%

\author{Nikolaos~Mitianoudis,~\IEEEmembership{Senior Member,~IEEE,}

\thanks{Manuscript received September 1, 2011; revised xx xx.}
\thanks{The author is with the Image Processing and Multimedia Laboratory, Department of Electrical and Computer
Engineering, Democritus University of Thrace, 67100 Xanthi, Greece (e-mail: nmitiano@ee.duth.gr, tel: +30 25410 79572, fax: +30 25410 79569.}}

%
%

\markboth{IEEE Transactions on Audio, Speech and Language Processing,~Vol.~x, No.~x, January~20xx}%
{MITIANOUDIS\MakeLowercase{\textit{}}: A Generalised Directional Laplacian Distribution: Characterisation, Estimation, Mixture Models and Applications}
%



\maketitle

\begin{abstract}

Directional or Circular statistics are pertaining to the analysis and interpretation of directions or rotations. In this work, a novel probability distribution is proposed to model multidimensional sparse directional data. The Generalised Directional Laplacian  Distribution (DLD) is a hybrid between the Laplacian distribution and the von Mises-Fisher distribution. The distribution's parameters are estimated using Maximum-Likelihood  Estimation over a set of training data points. Mixtures of Directional Laplacian Distributions (MDLD) are also introduced in order to model multiple concentrations of sparse directional data.  The author explores the application of the derived DLD mixture model to cluster sound sources that exist in an underdetermined instantaneous sound mixture. The  proposed model can solve the general $K\times L$ ($K<L$) underdetermined instantaneous source separation problem, offering a fast and stable solution.

\end{abstract}


\ifCLASSOPTIONpeerreview
\begin{center}\bfseries EDICS Category:  AUD-SSEN \end{center}
\fi
\IEEEpeerreviewmaketitle

\theoremstyle{definition}
\newtheorem{defin}{Definition}

\section{Introduction}
\IEEEPARstart{A}{ngles}, rotations, months and days fall into the same category commonly known as {\em circular} or  {\em directional} data, since they can be represented by points on the surface of the unit $p$-dimensional sphere~\cite{CircularStats}. {\em Circular Statistics} is the branch of statistics that addresses the modeling and inference from circular data, i.e. data with rotating values. To model directional data, one can generate many interesting circular models from known probability distributions by either wrapping a linear distribution around the unit circle or transforming a bivariate linear r.v. to its directional component~\cite{CircularStats}. However, there exist distributions that are periodic by definition and can therefore offer closed-form models for circular or directional data.

The {\em von Mises distribution} (also known as the circular normal distribution) is a continuous probability distribution on the unit circle~\cite{CircularStats,Fisher}. It may be considered the circular equivalent of the normal distribution and is defined by:
\begin{equation}
p(\theta)=\frac{e^{k\cos(\theta-m)}}{2\pi I_0(k)}\quad, \forall\textrm{ }\theta\in[0,2\pi)
\label{vonMises1}
\end{equation}
where $I_0(k)$ is the modified Bessel function of the first kind of order $0$, $m$ is the mean and $k>0$ describes the ``width'' of the distribution. Recently, Gattoa and Jammalamadaka~\cite{Gattoa07} proposed a ``Generalized von Mises'' (GvM) distribution in the form of $p(\theta)\propto e^{k_1 \cos(\theta-m_1) + k_2 \sin(\theta-m_2)}$, offering symmetric, asymmetric, unimodal or bimodal varieties of the original von Mises distribution.

A generalisation of the previous density  is the $p$-D  von Mises-Fisher distribution~\cite{Dhillon03,Mardia}. A $p$-dimensional unit random vector $\mathbf{x}$ $(||\mathbf{x}||=1)$ follows a {\em von Mises-Fisher} distribution, if its probability density function is described by:
\begin{equation}
p(\mathbf{x})\propto e^{k\mathbf{m}^T\mathbf{x}}\quad, \forall ||\mathbf{x}||\in \mathcal{S}^{p-1}
\label{NDMises}
\end{equation}
where $||\mathbf{m}||=1$ defines the centre, $k\geq 0$ and $\mathcal{S}^{p-1}$ is the $p$ dimensional unit hypersphere. Since the random vector $\mathbf{x}$ resides on the surface of a $p$-D unit-sphere, $\mathbf{x}$ essentially describes directional data. In the case of $p=2$,  $\mathbf{x}$ models data that exist on the unit circle and thus can be described only by an angle. In this case, the von Mises-Fisher distribution is reduced to the von-Mises distribution of (\ref{vonMises1}). The von Mises-Fisher distribution has been extensively studied and many methods have been proposed to fit the distribution or its mixtures to normally distributed circular data~\cite{Bentley06,CircularStats, Mardia,Dhillon03}.

This study proposes a novel distribution to model directional sparse data. {\em Sparsity} is mainly used to describe data that are mostly close to their mean value with the exception of several outlying values. There are several sparse models that have been proposed for linear sparse data~\cite{Lewicki99NC}. The Laplacian distribution $p(x)\propto e^{k|x-m|}$ appears to be a strong candidate in modelling sparse data~\cite{Davies00,Lewicki99NC}. In \cite{Eltoft06}, Eltoft et al proposed a multidimensional extension of the Laplacian distribution for $p$-D random variables with infinite support and provided parameter estimation algorithms for the proposed distribution and its mixtures. In \cite{Kotz00}, Kotz et al provided a multidimensional asymmetric model for the Laplacian distribution, which is a generalization of the previous approach again for $p$-D random variables with infinite support.  There were several attempts to model circular sparse signals by wrapping an 1-D or multidimensional Laplace distributions of infinite support~\cite{Jammalamadaka,Ghosh04,Mitianoudis07c}. The density wrapping solution is reported to have increased computational cost, as it is equivalent to using mixture models (the periodic repetition of a density function is equivalent to a mixture of density functions)~\cite{Mitianoudis07f}. Building from the original von Mises-Fisher distribution, this work proposes a Generalised Directional  Laplacian Distribution (DLD) as a direct modelling solution for multidimensional directional sparse data. The Maximum Likelihood  estimates (MLE) of the model's parameters are derived, along with an Expectation-Maximisation (EM) algorithm that estimates the parameters of a Mixture of Directional Laplacian Distributions (MDLD).

One application where directional statistical modelling is essential is {\em Underdetermined Audio Blind Source Separation} (BSS)~\cite{Lewicki99NC,Hyva98NC,Zibulevsky02,Grady04b, Mitianoudis07c,Mitianoudis07f}. Assume that a set of $K$ sensors $\mathbf{x}(n)=[x_1(n),\dots,x_K(n)]^T$ observes a set of $L$ $(K<L)$ sound sources $\mathbf{s}(n)=[s_1(n),\dots,s_L(n)]^T$. The instantaneous (anechoic) mixing model can  be expressed in mathematical terms, by
\begin{equation}
\mathbf{x}(n)=\mathbf{A}\mathbf{s}(n)
\end{equation}
where $\mathbf{A}$ represents a $K\times L$ {\em mixing matrix} and $n$ the sample index. Blind source separation algorithms provide an estimate of the source signals $\mathbf{s}$ and the mixing matrix $\mathbf{A}$, based on the observed microphone signals and some general statistical source profile. A variety of solutions exist for the complete instantaneous case ($K=L$) providing hiqh-quality separation (for more information, please refer to \cite{ICAbook01,Cichocki02,ICAbook02}). The underdetermined instantaneous case is more challenging, since in this case, the estimation of the mixing matrix $\mathbf{A}$ is not sufficient for the estimation of the source signals $\mathbf{s}$~\cite{Mitianoudis07c}. The two-channel ($K=2$) BSS scenario has been examined in detail in the past~\cite{Lewicki99NC,Hyva98NC,Zibulevsky02,Grady04b,Mitianoudis07c}. In this particular case, the source separation problem is reduced to an angular clustering problem of sparse data, as initially introduced by Hyv\"arinen~\cite{Hyva98NC} and Zibulevsky et al~\cite{Zibulevsky02}. O'Grady and Pearlmutter~\cite{Grady04} proposed an algorithm to perform separation via Oriented Lines Separation (LOST) using clustering along lines in a similar manner to Hyv\"arinen~\cite{Hyva98NC}. Davies and Mitianoudis~\cite{Davies02b} employed two-state Gaussian Mixture Models (GMM) to model the source densities in a sparse representation and also the possible additive noise. In~\cite{Mitianoudis07f}, the authors introduced Laplacian Mixture Models to perform angular clustering of sparse sources. To tackle the angular wrapping at $\pi$, the authors also examined the use of Wrapped Laplacian Mixtures (MoWL)~\cite{Mitianoudis07c}. However, the last two efforts do not offer a closed form solution to the problem and they can not be easily expanded to more than two sensors. Recently, Arberet et al~\cite{Arberet10} proposed a method to count and locate sources in underdetermined mixtures. Their approach is based on the hypothesis that in localised neighbourhoods around some time-frequency points $(t,f)$ (in the Short-Time Fourier Transform (STFT) representation) only one source essentially contributes to the mixture. Thus, they estimate the most dominant source (the Estimated Steering Vector) and a local confidence Measure which increases where a single component is only present. A clustering approach merges the above information and estimates the mixing matrix $\mathbf{A}$. In~\cite{Vincent09}, Vincent et al used local Gaussian Modelling of minimal constrained variance of the local time-frequency neighbours assuming knowledge of the mixing matrix $\mathbf{A}$. The candidate sources' variances are estimated after minimising the Kullback-Leibler (KL) divergence between the empirical and expected mixture covariances, assuming that at maximum 3 sources contribute to each time-frequency neighbourhood  and the sources are derived using Wiener filtering. There are also a number of source separation approaches that attempt to solve the convolutive underdetermined source separation problem. In this setup, the sound sources are recorded in a room and the elements of the mixing matrix $\mathbf{A}$ are replaced by FIR filters modelling the impulse responses between each source and microphone. Sawada et al~\cite{Sawada07,Araki07,Sawada11}, Winter et al~\cite{Winter07}, Duong et al~\cite{Vincent10} and many other researchers have proposed a variety of algorithms that can tackle the convolutive mixture problem; however, these approaches go beyond the scope of this paper, which is instantaneous underdetermined source separation.

This study extends previous work by Mitianoudis and Stathaki~\cite{Mitianoudis07c,Mitianoudis07f}. The proposed multidimensional DLD model offers a closed form solution to the modelling of directional sparse data and can also address the general $K\times L$ underdetermined source separation problem, which is rarely tackled in the literature. In addition, the proposed model is more computationally efficient compared to the warped laplacian solution in~\cite{Mitianoudis07f}.  The derived MLE algorithms are tested with several synthetic modelling experiments and real audio BSS examples and are compared with the solution of Vincent et al~\cite{Vincent09} that can address the general multichannel problem.

\section{A Generalised Directional Laplacian model}
\subsection{Definition}

Assume a r.v. $\theta$ modelling directional data with $\pi$-periodicity. The periodicity of the density function can be amended to reflect a ``fully circular'' phenomenon ($2\pi$), however, for the rest of the paper we will assume that $\theta\in [0,\pi)$, since it is required by the source separation application. From the definition of the von-Mises distribution in (\ref{vonMises1}), one can create a Laplacian structure simply by introducing a $|\cdot |$ operator in the superscript of the exponential. This action introduces a large concentration around the mean, which is needed to describe a sparse or Laplacian density. Values far away from the mean are smoothed out by the exponential. Additionally, we have to perform some minor amendments to the phase shift and also invert the distribution in order to impose the desired shape on the derived density.
\begin{defin}
The following probability density function models directional Laplacian data over $[0,\pi)$ and is termed {\em Directional Laplacian Density} (DLD):
\begin{equation}
p(\theta)=c(k) e^{-k|\sin(\theta-m)|}\quad, \forall\textrm{ }\theta\in[0,\pi)
\label{DLD}
\end{equation}
where $m\in[0,\pi)$ defines the mean, $k>0$ defines the width (``approximate variance'') of the distribution, $c(k)=\frac{1}{\pi I_0(k) }$ and $I_0(k)=\frac{1}{\pi}\int_0^\pi e^{-k\sin\theta}d\theta$.
\end{defin}

The normalisation coefficient $c(k)=1/\pi I_0(k)$ is derived from the fundamental normalisation property of probability density functions~\cite{Mitianoudis10a}. Examples of ({\ref{DLD}}) and more details on the special 1D DLD case can be found in \cite{Mitianoudis10a}.

The next step is to derive a generalised definition for the Directional Laplacian model. To generalise the concept of 1D DLD in the $p$-dimensional space, we will be inspired by the $p$-D  von Mises-Fisher distribution~\cite{Dhillon03,Mardia}. The von Mises-Fisher distribution is described by $p(\mathbf{x})\propto e^{k\mathbf{m}^T\mathbf{x}}$ (see (\ref{NDMises})). Since $||\mathbf{x}||=||\mathbf{m}||=1$, the inner product $\mathbf{m}^T\mathbf{x}=\cos\psi$, where $\psi$ is the angle between the two vectors $\mathbf{x}$ and $\mathbf{m}$. Following a similar methodology to the 1D-DLD, we need to formulate the term $-k|\sin\psi|$ in the superscript of the exponential. It is straightfoward to derive $|\sin\psi| =\sqrt{1 -\cos^2\psi}=\sqrt{1 -(\mathbf{m}^T\mathbf{x})^2}$. Thus, the superscript of the generalised DLD can be given by $-k\sqrt{1 -(\mathbf{m}^T\mathbf{x})^2}$.

\begin{defin}
The following probability density function models $p$-D  directional Laplacian data and is termed {\em Generalised Directional Laplacian Distribution} (DLD):
\begin{equation}
p(\mathbf{x})=c_p(k)e^{-k\sqrt{1-(\mathbf{m}^T\mathbf{x})^2}}\quad, \forall\textrm{ }||\mathbf{x}||\in \mathcal{S}^{p-1}
\label{MDDLD}
\end{equation}
where $\mathbf{m}$ defines the mean, $k\geq 0$ defines the width (``approximate variance'') of the distribution, $c_p(k)=\frac{\Gamma(\frac{p-1}{2})}{\pi^{\frac{p+1}{2}} I_{p-2}(k)}$, $I_{p}(k)=\frac{1}{\pi}\int_{0}^{\pi} e^{-k\sin\theta}\sin^{p}\theta d\theta$ and $\Gamma(\cdot)$ represents the Gamma function\footnote{Note that for $n$ positive integer, we have that $\Gamma(n)=(n-1)!$}.
\end{defin}

\begin{figure}
\centering
\includegraphics[width=3.5in]{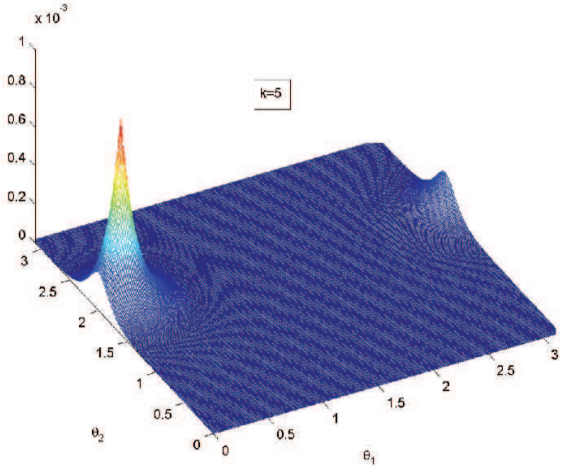}
\caption{The proposed Generalised Directional Laplacian Distribution for $k=5$ and $p=3$.}
\label{fig4}
\end{figure}

The normalisation coefficient $c_p(k)$ is calculated in Appendix \ref{appxMD1}. In the case of $p=2$, the  generalised DLD is reduced to the one dimensional DLD of (\ref{DLD}), verifying the validity of the above model. The generalised DLD density models ``directional'' data on the half-unit $p$-D sphere, however, it can be extended to the unit $p$-D sphere, depending on the specifications of the application. In Figure \ref{fig4}, an example of the generalised DLD is depicted for $p=3$ and $k=5$. The centre $\mathbf{m}$ is calculated using spherical coordinates $\mathbf{m}=[\cos\theta_1\cos\theta_2;\textrm{ } \cos\theta_1\sin\theta_2;\textrm{ }\sin\theta_1]$ for $\theta_1=0.2$ and $\theta_2=2$.

\subsection{Generalised Directional Laplacian Density samples generation}
To generate 1D Directional Laplacian data, we employed the inversion of the cumulative distribution method~\cite{Devroye86}. Inversion methods are based on the observation that continuous cumulative distribution functions (cdf) range uniformly over the interval $(0, 1)$. Since the proposed density is bound between $[0, \pi)$, we can evaluate the cdf of the Directional Laplacian density with uniform sampling at $[0, \pi)$ and approximate the inverse mapping using spline interpolation. Thus, uniform random data in the interval $(0, 1)$ can be transformed to 1D Directional Laplacian random samples, using the described inverse mapping procedure.

To simulate $2$-D Directional Laplacian random data ($p=3$), we sampled the 2-D density function for specific $\mathbf{m}$, $k$. The bounded value space $(\theta_1,\theta_2\in [0,\pi))$ is quantised into small rectangular blocks, where the density is assumed to be uniform. Consequently, we generate a number of uniform random samples for each block. The number of samples generated from each block is different and defined by the overall DL density. The required $3$-D unit-norm random vectors are produced using spherical coordinates with unit distance and angles $\theta_1, \theta_2$ from the random 2-D Directional data. The above procedure can be extended for the generation of $p$-D directional data.

\subsection{Maximum Likelihood Estimation of parameters $\mathbf{m}$, $k$}
\begin{figure}
\centering
\includegraphics[width=3in]{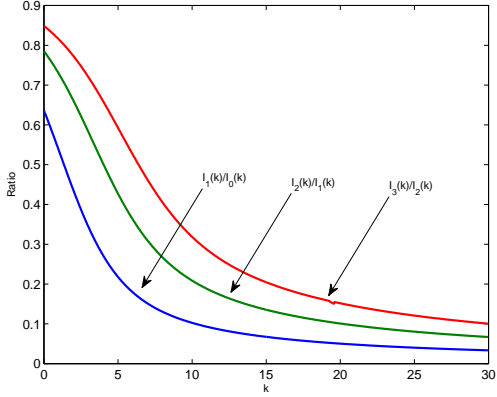}
\caption{The ratio $I_{p}(k)/I_{p-1}(k)$ is a monotonic $1-1$ function of $k$.}
\label{fig2}
\end{figure}

Assume a population of $p$-dimensional angular data $\mathbf{X}=\{\mathbf{x}_1,\dots,\mathbf{x}_n,\dots,\mathbf{x}_N\}$ that follow a $p$-dimensional Directional Laplacian Distribution. To estimate the model parameters using Maximum Likelihood Estimation (MLE), one can form the log-likelihood and estimate the parameters $\mathbf{m}$, $k$ that maximise it. For the Generalised DLD density, the log-likelihood function can be expressed, as follows:
\begin{equation}
J(\mathbf{X},\mathbf{m},k)=N\log \frac{\Gamma(\frac{p-1}{2})}{\pi^{\frac{p+1}{2}}I_{p-2}(k)} - k\sum_{n=1}^N \sqrt{1-(\mathbf{m}^T\mathbf{x}_n)^2}
\label{LogLike_MDDLD}
\end{equation}
Alternate optimisation is performed to estimate $\mathbf{m}$ and $k$. The gradients of $J$ along $\mathbf{m}$ and $k$ are calculated in Appendix \ref{appxMD2}. The update for $\mathbf{m}$ is given by gradient ascent on the log-likelihood via:
\begin{equation}
\mathbf{m}^+\leftarrow \mathbf{m} +\eta\sum_{n=1}^N\frac{\mathbf{m}^T\mathbf{x}_n}{\sqrt{1-(\mathbf{m}^T\mathbf{x}_n)^2}}\mathbf{x}_n
\label{estimM2}
\end{equation}
\begin{equation}
\mathbf{m}^+\leftarrow \mathbf{m}^+/||\mathbf{m}^+||
\label{estimM22}
\end{equation}
where $\eta$ defines the gradient step size. Since the gradient step does not guarantee that the new update for $\mathbf{m}$ will remain on the surface of $\mathcal{S}^{p-1}$, we normalise the new update to unit norm. To estimate $k$, a numerical solution to the  equation ${\partial J(\mathbf{X},\mathbf{m},k)}/{\partial k}=0$ is estimated. From the analysis in Appendix \ref{appxMD2}, we have that
\begin{equation}
\frac{I_{p-1}(k)}{I_{p-2}(k)}=\frac{1}{N}\sum_{n=1}^{N}\sqrt{1-(\mathbf{m}^T\mathbf{x}_n)^2}
\label{estimK2}
\end{equation}
To calculate $k$ analytically from the ratio $I_{p-1}(k)/I_{p-2}(k)$ is not straightforward. However, after numerical evaluation, it can be demonstrated that the ratio $I_{p-1}(k)/I_{p-2}(k)$ is a smooth monotonic $1-1$ function of $k$. In Figure \ref{fig2}, the ratio $I_{p-1}(k)/I_{p-2}(k)$ is estimated for uniformly sampled values of $k\in [0.01,30]$ and $p = 2, 3, 4$. Since this ratio is not dependent on data, one can create a look-up table for a variety of $k$ values and use interpolation to estimate $k$ from an arbitrary value of $I_{p-1}(k)/I_{p-2}(k)$. This look-up table solution is more efficient compared to possible iterative estimation approaches of $k$ and generally accelerates the model's training.

\subsection{Mixtures of Generalised Directional Laplacians}
\label{MMDDLD_EM}
 One can employ {\em Mixtures of Generalised Directional Laplacians} (MDLD) in order to model multiple concentrations of directional generalised ``heavy-tailed signals''.
\begin{defin}{\em Mixtures of Generalised Directional  Laplacian Distributions} are defined by the following pdf:
\begin{equation}
p(\mathbf{x})=\sum_{i=1}^K a_i c_p(k_i)  e^{-k_i\sqrt{1-(\mathbf{m}_i^T\mathbf{x})^2}}\quad, \forall\textrm{ }||\mathbf{x}||\in \mathcal{S}^{p-1}
\label{MMDDLD}
\end{equation}
where $a_i$ denotes the weight of each distribution in the mixture, $K$ the number of DLDs used in the mixture and $\mathbf{m}_i$, $k_i$ denote the mean and the ``width'' (approximate variance) of each distribution.
\end{defin}
The mixtures of DLD can be trained using the Expectation-Maximisation (EM) algorithm. Following the previous analysis in~\cite{Bilmes98,Mitianoudis07c,Mitianoudis07f}, one can yield the following simplified likelihood function:
\begin{equation}
 \mathcal{L}(a_i,\mathbf{m}_i,k_i)=
\end{equation}
\begin{equation}
\sum_{n=1}^N \sum_{i=1}^K\left(\log \frac{a_i \Gamma(\frac{p-1}{2})}{\pi^{\frac{p+1}{2}}I_{p-2}(k)} - k \sqrt{1-(\mathbf{m}^T\mathbf{x})^2}\right)p(i|\mathbf{x}_n)\nonumber
\end{equation}
where $p(i|\mathbf{x}_n)$ represents the probability of sample $\mathbf{x}_n$ belonging to the $i^{th}$ Directional Laplacian of the mixture. In a similar fashion to other mixture model estimation, the updates for $p(i|\mathbf{x}_n)$ and $\alpha_i$ can be given by the following equations:
\begin{equation}
p(i|\mathbf{x}_n)\leftarrow \frac{a_i c_p(k_i) e^{-k_i\sqrt{1-(\mathbf{m}_i^T\mathbf{x})^2}}}{\sum_{i=1}^Ka_i c_p(k_i) e^{-k_i\sqrt{1-(\mathbf{m}_i^T\mathbf{x})^2}}}
\end{equation}
\begin{equation}
a_i\leftarrow \frac{1}{N}\sum_{n=1}^Np(i|\mathbf{x}_n)
\end{equation}
Based on the derivatives calculated in Appendix \ref{appxMD2}, it is straightforward to derive the following updates for $\mathbf{m}_i$ and $k_i$, as follows:
\begin{equation}
\mathbf{m}_i^+\leftarrow \mathbf{m}_i +\eta\sum_{n=1}^N k_i\frac{\mathbf{m}^T\mathbf{x}_n}{\sqrt{1-(\mathbf{m}^T\mathbf{x}_n)^2}}\mathbf{x}_np(i|\mathbf{x}_n)
\label{estimM3}
\end{equation}
\begin{equation}
\mathbf{m}_i^+\leftarrow \mathbf{m}_i^+/||\mathbf{m}_i^+||
\end{equation}
To estimate $k_i$, we solve the equation ${\partial I}/{\partial k_i}=0$ numerically. The equation yields:
\begin{equation}
\frac{I_{p-1}(k_i)}{I_{p-2}(k_i)}=\frac{\sum_{n=1}^{N}\sqrt{1-(\mathbf{m}_i^T\mathbf{x}_n)^2}p(i|\mathbf{x}_n)}{\sum_{n=1}^{N}p(i|\mathbf{x}_n)}
\label{estimK3}
\end{equation}
The training of this mixture model is also dependent on the initialisation of its parameters, especially the means $\mathbf{m}_i$~\cite{Mitianoudis07c}. In Appendix \ref{appx4}, the standard K-Means algorithm is reformulated in order to tackle $p$-dimensional directional data. The proposed $p$-dimensional {\em Directional K-Means} is used to initialise the means $\mathbf{m}_i$ of the DLDs in the generalised DLD mixture EM training.  A {\em Directional K-Means} already exists in the literature~\cite{Banerjee05}, however, the proposed $p$-dimensional {\em Directional K-Means} in Appendix \ref{appx4} employs a distance function more relevant to sparse directional data.

\section{Audio Source Separation using Mixtures of DLD}

In underdetermined audio source separation a set of $K$ sensors $\mathbf{x}(n)=[x_1(n),\dots,x_K(n)]^T$ observes a set of $L$ $(K<L)$ sound sources $\mathbf{s}(n)=[s_1(n),\dots,s_L(n)]^T$. The instantaneous (anechoic) mixing model can  be expressed in mathematical terms, by $\mathbf{x}(n)=\mathbf{A}\mathbf{s}(n)$, where $\mathbf{A}$ represents a $K\times L$ {\em mixing matrix}. The underdetermined instantaneous source separation problems consists of two sub-problems a) estimate the mixing matrix $\mathbf{A}$, b)  estimate the sound sources $\mathbf{s}(n)$, given the observed signals $\mathbf{x}(n)$~\cite{Mitianoudis07f}. The solution of this problem can have a {\em unique} and {\em identifiable} solution, according to Eriksson and Koivunen~\cite{Eriksson03}, as long i) there are no Gaussian sources present in the mixture, ii) the mixing matrix $\mathbf{A}$ is of full row rank, i.e. $\textrm{rank}(\mathbf{A})=M$ and iii) none of the source variables has a characteristic function featuring a component in the form $\exp(Q(u))$, where $Q(u)$ is a polynomial of a degree of at least two.

Assume a two-sensor instantaneous mixing approach ($K=2$) and that the source signals $s_i(n)$ are sparse. When the sources are sparse, smaller coefficients are more probable, whereas all the signal's energy is concentrated in few large values. Therefore, the density of the data in the mixture space shows a tendency to cluster along the directions of the mixing matrix columns~\cite{Zibulevsky02}. That is to say, that the phase difference $\theta_n=\textrm{atan}\frac{x_2(n)}{x_1(n)}$ between the two sensors can be used to identify source concentrations (clusters). The centres of the clusters denote the columns of the mixing matrix~\cite{Mitianoudis07f}. Using the phase difference information between the two sensors is equivalent to mapping all the observed data points on the unit-circle. This is equivalent to the concept of mapping all the observed data points to the half-unit $p$-dimensional sphere, as proposed by Zibulevsky et al~\cite{Zibulevsky02}. Thus, the general underdetermined source separation problem becomes a directional clustering problem on the half-unit $p$-dimensional sphere. For a more detailed analysis of the above, the reader is referred to~\cite{Zibulevsky02,Grady04b,Mitianoudis07c,Mitianoudis07f,Arberet10,ICAbook02}.

In~\cite{Mitianoudis07c}, the authors introduced the concept of {\em Mixture of Laplacians} (MoL) to tackle this angular clustering problem in the case of a two-sensor setup. Once the MoL was fitted to the angular data $\theta_n$, each source was represented by each of the Laplacians in the mixture. Separation was performed either by hard thresholding or soft (fuzzy) thresholding. This solution suffered from clusters centred closer to $0^o$ or  $180^o$, since the Laplacian distribution used in these Mixture models has infinite instead of a circular support. To offer a more complete solution to this problem, in ~\cite{Mitianoudis07f}, the authors proposed a {\em Mixture of Warped Laplacians} (MoWL) (i.e. periodic repetitions of the Laplacian density) that tackles clustering across the borders. Neverless, the two approaches handled only the two-sensor case (1D) and the speed of training MoWL was rather slow, as it is equivalent to training two mixture models (one EM for the warping of each Laplacian and one EM for the mixture of warped Laplacians).

The generalised Directional Laplacian Density offers a faster and complete solution to the problem, since the proposed function addresses directional data by definition and is multidimensional, which implies that it can be automatically applied to the general $K\times L$ separation scenario. Once the Mixtures of DLD are fitted to the multichannel directional data, separation can be performed by ''hard-thresholding'' for the 1D-case (intersections of the individual DLDs), or ''soft-thresholding'' for the general p-D case in a similar manner to~\cite{Mitianoudis07f}. That is to say, we can attribute points that constitute a chosen ratio $q$ (i.e. $0.7-0.9$) of the density of each DLD to the corresponding source. Hence, the $i^{th}$ source can be associated with those points on the unit $x_n$ $p$-dimensional sphere , for which $p(x_n)\ge(1-q)\alpha_ic_p(k_i)$, where $p(x_n)$ is given by (\ref{MMDDLD}).

Having attributed the points $\mathbf{x}(n)$ to the $L$ sources, using either the ``hard'' or the ``soft'' thresholding technique, the next step is to reconstruct the sources. Let $S_i\sqsubseteq N$ represent the point indices that have been attributed to the $i^{th}$ source and $\mathbf{m}_i$ the corresponding mean vector, i.e. the corresponding column of the mixing matrix. We initialise $u_i(n)=0 , \forall$ $n=1,\dots,N$ and $i=1,\dots,L$. The source reconstruction is performed by substituting:
\begin{equation}
u_i(S_i)=\mathbf{m}_i^T\mathbf{x}(S_i)\qquad \forall\textrm{ }i=1,\dots,L
\end{equation}
In the case that we need to capture the multichannel image of the separated source, the result of the separation is a multichannel output that is initialised to $\mathbf{u_i}(n)=\mathbf{0}, \forall$ $n=1,\dots,N$. The source image reconstruction is performed by:
\begin{equation}
\mathbf{u_i}(S_i)=\mathbf{x}(S_i)\qquad \forall\textrm{ }i=1,\dots,L
\end{equation}
\section{Experiments}
In this section, we verify the validity of the above derived MLE algorithms and the goodness-of-fit of the proposed Directional Laplacian model and its mixtures. The first part of the evaluation process contains several synthetic examples that verify the principles of the derived algorithms. The second part demonstrates the density's relevance and performance  in underdetermined audio source separation. At this point, we need to clarify that the main scope of the paper is the proposal of a novel multi-dimensional density that can find applications in many other fields, including underdetermined source separation. Therefore, we are not aiming at proposing the best-performing source separation algorithm, but an algorithm that improves our previous efforts both in stability, speed and performance and offers a fast alternative to state-of-the-art algorithms with reasonable separation performance.

For the rest of the section, we note that the integral $I_{p}(k)$ was numerically estimated using MATLAB's $\texttt{quadl}$ command. As mentioned earlier, the estimation of $k$ from equations (\ref{estimK2}), (\ref{estimK3}) is performed using spline interpolation (as implemented by MATLAB's $\texttt{interp1}$ command) from a look-up table for several values of $k$ and $I_{p-1}(k)/I_{p-2}(k)$  that is created and stored before optimisation.

\begin{figure}
\centering
\subfigure[MLE for a DLD]{\includegraphics[width=1.7in]{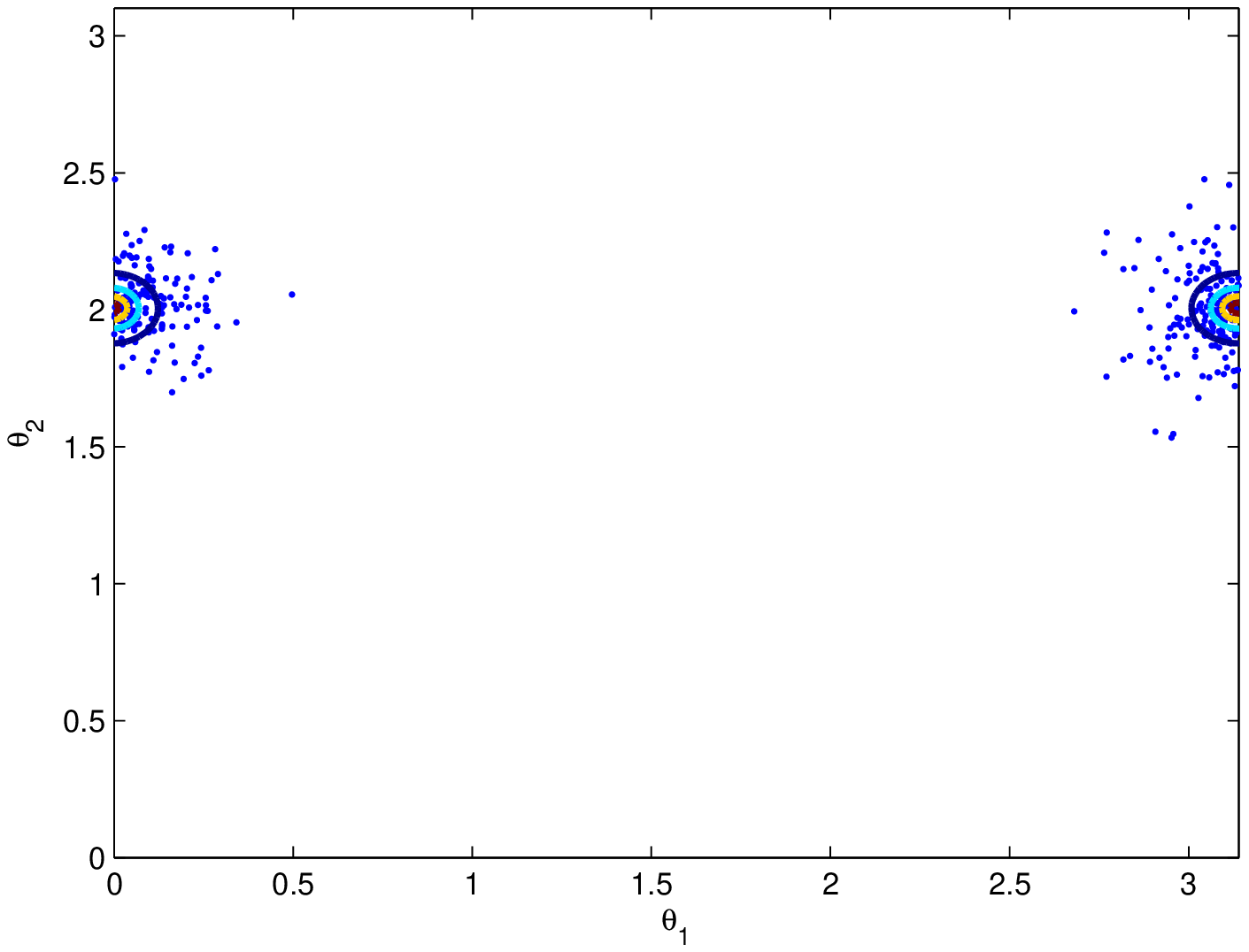}}
\subfigure[EM estimation for a DLD mixture]{\includegraphics[width=1.7in]{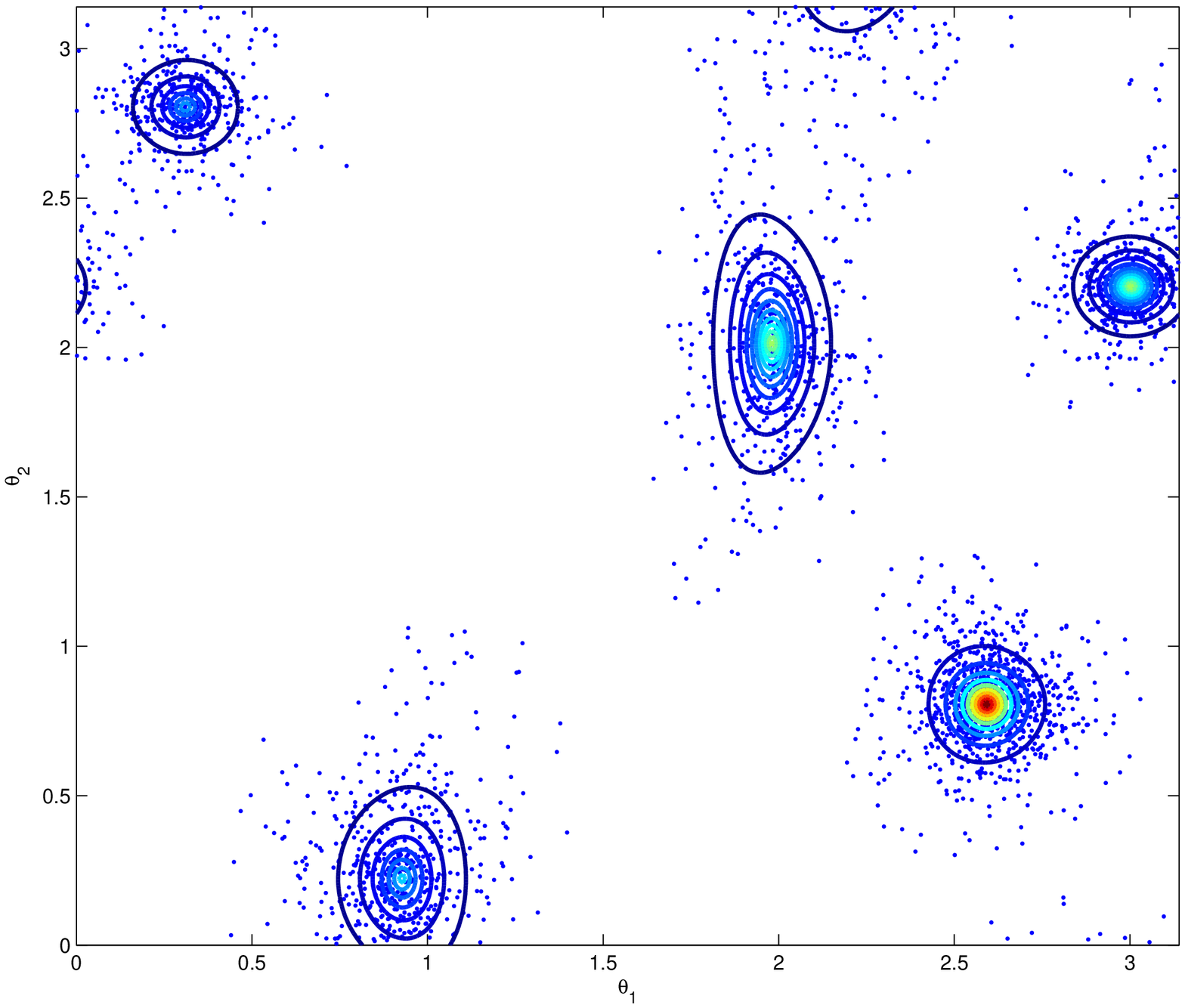}}
\caption{Examples of 2D ML parameter estimation for the DLD model (left) and its mixture model (right) using $2000$ randomly generated 2D Directional Laplacian data.}
\label{fig5.5}
\end{figure}

\subsection{Synthetic Examples}
\begin{figure}
\centering
\includegraphics[width=2.7in]{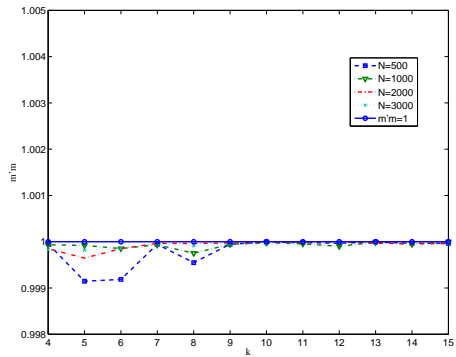}
\caption{Estimation of  $\mathbf{m}^T\hat{\mathbf{m}}$ for various values of $N=500, 1000, 2000, 3000$ and $k\in[4, 15]$ for the $p=3$ (2D case).}
\label{fig5.8}
\end{figure}

The first step was to test the derived algorithms with synthetic data. The 2-dimensional case ($p=3$) was selected in order to facilitate the visualisation of the training results. We explored various cases of $\mathbf{m}, k , N$, especially centres that are closer to the wrapping boundaries of $0$ and $\pi$. For the MLE of the DLD's parameters, we employed equations (\ref{estimM2}), (\ref{estimM22}) and (\ref{estimK2}) with random initialisation of the centres and $k$. The gradient step size value of $\eta=0.01$ in (\ref{estimM2}) (and in (\ref{estimM3})) has shown efficient and fast convergence for all the experiments in the paper. Similarly to all gradient-based iterative optimisation algorithms, a ``bad'' choice of $\eta$ may lead to either slow convergence or inaccurate optimum estimation. Keeping the $\eta=0.01$ did not seem to affect the estimation performance in our experiments.  In Figure \ref{fig5.5}(a), an example of fitting the 2D DLD on $2000$ directional Laplacian samples centred close to the wrapping border and $k=15$ is presented. In this figure, the data-point scatter plot is overlaid by a contour plot of the fitted 2D-DLD model. To evaluate the efficiency of $\mathbf{m}$ estimation, we examined several extreme cases summarised in Table \ref{Table2}. For each different experiment, we evaluated $50$ independent runs  with random directional Laplacian data. The average estimates of $\mathbf{m}^T\mathbf{\hat{m}}$  for each case are displayed in Table \ref{Table2}. It is evident that one can get very accurate results in terms of $\mathbf{\hat{m}}$ (estimate of $\mathbf{m}$), regardless of the dataset size $N$ for fairly concentrated data (values of $k>6$). The effect of sample size $N$ is also demonstrated in Figure \ref{fig5.8}. The estimation of $\mathbf{m}^T\hat{\mathbf{m}}$ for the 2D case is examined for values of $N=500, 1000, 2000, 3000$ and $k\in[4, 15]$. We can see that the estimate $\hat{\mathbf{m}}$ gets closer or identical to $\mathbf{m}$ for greater values of $k$ (i.e. more concentrated centres) and more data points.
\begin{table}
\caption{MLE of $\mathbf{m}$ for the 2D Directional Laplacian ($p=3$) for various values of $\mathbf{m}, k, N$. Average results for $50$ independent runs for each experiment.}
\begin{center}
{
\begin{tabular}[width=0.1\textwidth]{|c|c|c|c|}\hline
$\mathbf{m}$ & $k$& $N$ & $\mathbf{m}^T\mathbf{\hat{m}}$  \\\hline\hline
[-0.4329 0.3234 0.8415]&12&100&0.9994\\\hline
[-0.4329 0.3234 0.8415]&12&1000&1.000\\\hline
[-0.4329 0.3234 0.8415]&12&2000&1.000\\\hline
[-0.4329 0.3234 0.8415]&4&1000&0.9998\\\hline
[-0.4161 0 0.9093]&8&100&0.9995\\\hline
[-0.4161 0 0.9093]&8&1000&0.9999\\\hline
[-0.4161 0 0.9093]&15&100&0.9997\\\hline
[-0.4161 0 0.9093]&15&1000&1.0000\\\hline
[-0.4161  0.9093 0]&8&100&0.9994\\\hline
[-0.4161  0.9093 0]&8&1000&0.9999\\\hline
[-0.4161  0.9093 0]&15&100&0.9999\\\hline
[-0.4161  0.9093 0]&15&1000&1.000\\\hline
\end{tabular}
}
\end{center}
\label{Table2}
\end{table}

\begin{figure}
\centering
\subfigure[1D case ($p=2$)]{\includegraphics[width=2.7in]{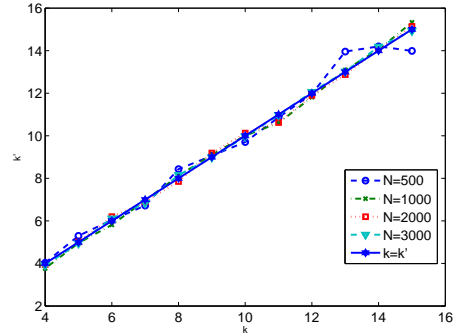}}
\subfigure[2D case ($p=3$)]{\includegraphics[width=2.7in]{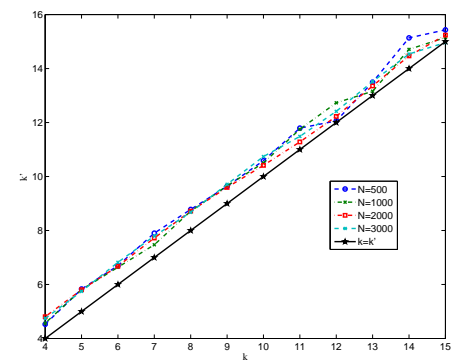}}
\caption{Estimation of $k$ for various values of $N=500, 1000, 2000, 3000$ and $k\in[4, 15]$ for $p=2$ (1D case) (a) and for $p=3$ (2D case) (b).}
\label{fig6}
\end{figure}

To evaluate the efficiency of $k$ estimation, we conducted a series of experiments for $N=500, 1000, 2000, 3000$ and $k\in[4,15]$. For each set of values $N, k$, we averaged the results of $50$ independent runs. The results are depicted in Figure \ref{fig6} for the 1D ($p=2$) and the 2D ($p=3$) case. The results demonstrate accurate estimates for all cases, especially for the 1D case. The estimation of $k$ for the 2D case seems to improve with the sample size, while the small difference between the estimated and the actual value of $k$ for small values of $k$ is due to possible model overfitting especially for smaller number of data points. This difference is very small and does not introduce any serious side-effects  in applications, such as audio source separation.

The next step is to evaluate the efficiency of the derived EM algorithm for the estimation of the $p$-D Directional Laplacian Density Mixtures.  We created a mixture of $5$ concentrations of 2D-DLD samples centred at various positions $\mathbf{m}_i$ and various values of $k_i$ and $a_i$, as summarised in Table \ref{Table4}. The total number of samples were $3000$. For the initialisation of the centres, we used the Directional K-Means algorithm, as described in Appendix \ref{appx4}. We ran $50$ independent runs of the EM-algorithm as described in Section \ref{MMDDLD_EM}. The average estimated $\mathbf{m}_i^T\mathbf{\hat{m}_i}$, $\hat{k}_i$ and $\hat{a}_i$ are depicted in Table \ref{Table4}.  We witnessed several incorrect initialisations caused by the Circular K-Means algorithm, especially in the smaller clusters (small $a_i$) or closely spaced clusters (around 7/50 times for DLD$_2$, whereas 0/50 times for DLD$_3$ or DLD$_5$). These incorrect initialisations resulted into a drop of the average performance. In the case of accurate initialisation, the clustering performance was very good. In Figure \ref{fig5.5} (right), we demonstrated a successful clustering and training of the DLD mixture for the synthetic source compilation. The random samples are depicted in a 2-D cluster plot along with the fitted MDLDs of the mixture. The clustering produced by the proposed EM algorithms seems to offer adequate accuracy.

\begin{table*}
\caption{Parameter Estimation for a Mixture of 2D-Directional Laplacian ($K=5, p=3$) using the proposed EM algorithm.  Average  parameter results for $50$ independent runs. }
\begin{center}
{
\begin{tabular}[width=0.1\textwidth]{|c||c|c|c|c|c|c|}\hline
&$\mathbf{m}_i$ & $k_i$& $a_i$ & $|\mathbf{m}_i^T\mathbf{\hat{m}}_i|$ & $|\hat{k_i}-k_i|/k_i$&$\hat{a_i}$\\\hline\hline
DLD$_1$&[-0.9001  0.3200 0.2955]&12&0.1333&0.9277&0.0431&0.1215\\\hline
DLD$_2$&[ 0.6092 0.1235 0.7833]&10&0.2&0.8730&0.1072&0.1663\\\hline
DLD$_3$&[-0.5970 -0.6147 0.5155]&14&0.3333&0.9997&0.0299&0.3259\\\hline
DLD$_4$&[0.1732 -0.3784 0.9093]&15&0.1667&0.98986&0.0986&0.2001\\\hline
DLD$_5$&[0.5826 -0.8004 0.1411]&15&0.1667&0.9995&0.0248&0.1779\\\hline
\end{tabular}
}
\end{center}
\label{Table4}
\end{table*}

\begin{figure}
\centering
\includegraphics[width=3.5in]{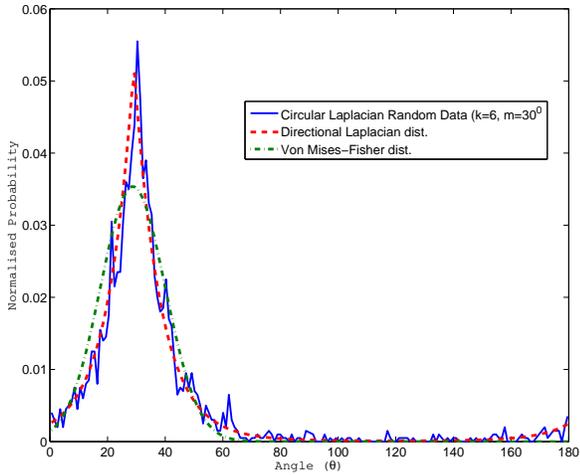}
\caption{Model fitting comparison between the DLD and the von Mises-Fisher distribution to Directional Laplacian Data ($m=30^o$, $k=6$).}
\label{fig7}
\end{figure}

\begin{figure}
\centering
\subfigure[1D case]{\includegraphics[width=3in]{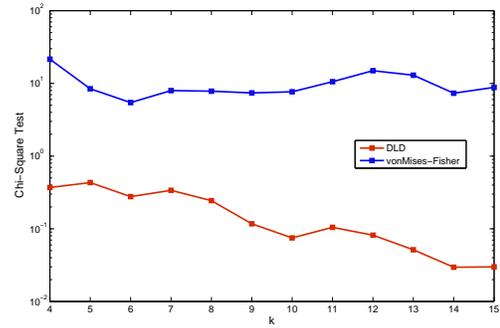}}
\subfigure[2D case]{\includegraphics[width=3in]{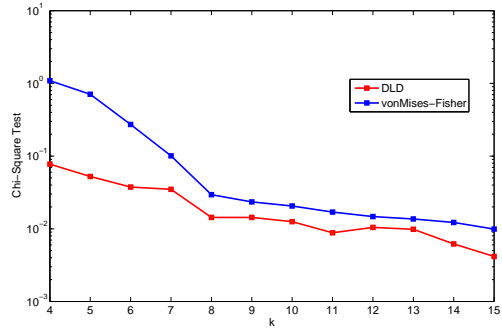}}
\caption{The Pearson Chi-Square Tests for the DLD and the von Mises-Fisher distribution for $k\in[4,15]$ and the 1D and 2D cases. The proposed DLD offers a closer fit to Laplacian data compared to the von Mishes-Fisher distribution.}
\label{fig8}
\end{figure}

\begin{figure}
\centering
\includegraphics[width=3in]{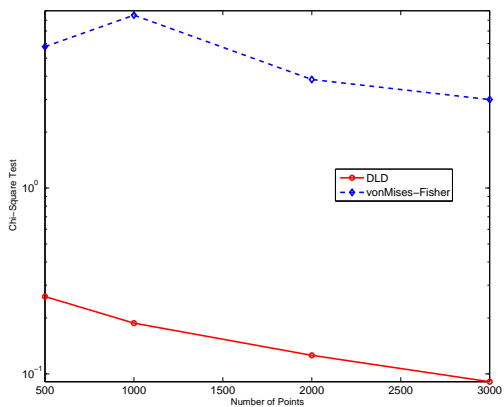}
\caption{The Pearson Chi-Square Tests for the DLD and the von Mises-Fisher distribution for $k=6$ for the 1D case as a function of number of functions. The goodness-of-fit increases with the number of training points for both distributions.}
\label{fig8.5}
\end{figure}

Finally, in order to compare the goodness-of-fit of the proposed DLD model with the von Mises-Fisher distribution, we generated $2000$ random Directional Laplacian 1D and 2D data for various values of $k$. Then, the proposed DLD MLE algorithm  and a von Mises-Fisher MLE algorithm~\cite{Dhillon03,Mardia} were used to fit the models to the data. An example of the two models fitted to the data is depicted in Figure \ref{fig7}. It can be observed that the proposed density offers a closer fit compared to the vonMises-Fisher density.  The Pearson Chi-Square test was calculated to compare the data normalised histogram with the fitted models~\cite{Huber02}. A lower Chi-Square score indicates a closer match of the fitted model to the actual data.  A comparison of the Pearson Chi-Square score for the two distributions for the 1D and the 2D case is depicted in Figure \ref{fig8}. It is clear that the proposed DLD model offers a closer match to the actual sparse data distribution compared to the more Gaussian-like von Mises-Fisher model. This conclusion applies for various values of $k$. A comparison  of the Pearson Chi-Square scores for $N=500, 1000, 2000, 3000$ points for the 1D case and $k=6$ is shown in Figure \ref{fig8.5}. The goodness-of-fit increases with the number of training points for both distributions. Since the proposed MDLD offers a closer fit for sparse data, it is rational to be preferred instead of the vonMises-Fisher to perform separation of sparse clusterings.

\begin{table*}[htb]
\caption{The proposed MDLD approach is compared for source estimation performance ($K=2$) in terms of SDR (dB), SIR (dB) and SAR(dB) with GaussSep, WMoL and Hyv\"arinen's approach. The measurements are averaged for all sources of each experiment.}
\begin{center}
{
\begin{tabular}[width=0.0.5\textwidth]{|l||c|c|c|c||c|c|c|c||c|c|c|c|}\hline
&\multicolumn{4}{c||}{\bf{SDR} (dB)}&\multicolumn{4}{c||}{\bf{SIR} (dB)}&\multicolumn{4}{c|}{\bf{SAR} (dB)}\\\hline
&MDLD&GaussSep&MoWL&Hyva&MDLD&GaussSep&MoWL&Hyva&MDLD&GaussSep&MoWL&Hyva\\\hline\hline
Latino1&6.38&5.51&5.72&0.89& 18.63&	8.96&	18.59&	9.61&6.93&9.20&6.26&3.63\\\hline
Latino2&3.21&4.71&2.10&	0.89& 11.50&8.87&	11.28&	9.61&4.95&	9.20&	3.85&3.63\\\hline
Groove&0.22&0.39&-0.43&-0.08&9.48&3.62&9.60&8.88&2.12&7.37&1.00&1.83\\\hline
Dev2Male3&3.04&6.22&	2.11&	-3.10& 13.69&	12.14&	13.30&	4.73&4.10&8.04&3.33&	2.72\\\hline
Dev2Female3&4.68&5.70&3.86&	-1.85& 15.28&	11.45&	16.58&	5.02&5.41&	7.51&4.61&	3.13\\\hline
Dev2WDrums&9.59&16.57&10.16&0.63& 19.77&23.83&19.98&7.57&10.55&	17.68&10.54&5.54\\\hline
Dev1WDrums&4.96&16.54&3.81&6.86& 13.88&20.94&	12.38&	16.75&6.37&	19.30&	5.20&	7.73\\\hline\hline
\it{Average}&4.58	&\bf{7.96}	&3.91&	0.6& \bf{14.61}&	12.83	&13.82&	8.88& 5.78&	\bf{11.19}	&4.97&	4.03\\\hline
\end{tabular}
}
\end{center}
\label{Table5}
\end{table*}

\subsection{Audio Source Separation}

In this section, we evaluate the proposed MDLD algorithm  for audio source separation.

We will use Hyv\"arinen's clustering approach~\cite{Hyva98NC}, the MoWL algorithm~\cite{Mitianoudis07c} and the ``GaussSep'' algorithm~\cite{Vincent09} for comparison. After fitting the MDLD with the proposed EM algorithm, separation will be performed using hard or soft thresholding, as described in our previous work~\cite{Mitianoudis07c,Mitianoudis07f}. In order to quantify the performance of the algorithms, we estimate the {\em Signal-to-Distortion Ratio} (SDR), the {\em Signal-to-Interference Ratio} (SIR) and the {\em Signal-to-Artifact Ratio} from the BSS$\_$EVAL Toolbox v.3~\cite{BSSeval}. The input signals for the MDLD, MoWL and Hyv\"arinen's approaches are sparsified using the {\em Modified Discrete Cosine Transformation} (MDCT), as developed by Daudet and Sandler~\cite{Daudet04}. The frame length for the MDCT analysis is set to $32$ msec for the speech signals and $128$ msec for the music signals sampled at $16$ KHz, and to $46.4$ msec for the music signals at $44.1$ KHz. We initialise the parameters of the MoWL and MDLD as follows: $\alpha_i=1/N$ and $c_i=0.001$,  $T=[-1,0,1]$ (for MoWL only) and $k_i=15$ (for the DLD only). The centres $m_i$ were initialised in either case using the Directional {\em K-means} step, as described in Appendix \ref{appx4}.  We used the ``GaussSep'' algorithm, as publicly available by the authors\footnote{MATLAB code for the ``GaussSep'' algorithm is available from {\tt http://www.irisa.fr/metiss/members/evincent/software}.}. For the estimation of the mixing matrix, we used Arberet et al's~\cite{Arberet10} DEMIX algorithm\footnote{MATLAB code for the ``DEMIX'' algorithm is available from {\tt http://infoscience.epfl.ch/record/165878/files/}.}, as suggested in \cite{Vincent09}. The number of sources in the mixture was also provided to the DEMIX algorithm, as it was provided to all other algorithms. The ``GaussSep'' algorithm operates in the STFT domain, where we used the same frame length with the other approaches and  a time-frequency neighbourhood size of $5$ for speech sources and $15$ for music sources.

\subsubsection{Two-microphone examples}
We tested the algorithms with the {\em Groove}, {\em Latino1} and {\em Latino2} datasets, available by BASS-dB~\cite{BASS-dB}, and sampled at $44.1$ KHz. The ``Groove'' dataset features four widely spaced sources: bass (far left), distorted guitar (center left), clean guitar (center right) and drums (far right). The two ``Latino'' datasets features four widely spaced sources: bass (far left), drums (center left), keyboards (center right) and distorted guitar (far right). We also used a variety of test signals from the Signal Separation Evaluation Campaigns SiSEC2008~\cite{SiSEC2008} and SiSEC2010~\cite{SiSEC2010}. We employed two audio instantaneous mixtures from the ``dev1'' and ``dev2'' data sets (``Dev2WDrums'' and  ``Dev1WDrums'' sets - 3 instruments at 16KHz) and two speech instantaneous mixtures from the ``dev2'' data set (``Dev2Male3'' and ``Dev2Female3'' sets - 4 closely located sources at 16 KHz). We used the development (dev) datasets instead of the test data sets, in order to have all the source audio files for proper benchmarking.

In Table \ref{Table5}, we can see the results for the four methods in terms of SDR, SIR and SAR. For simplicity, we averaged the results for all sources at each experiment. The reader of the paper can visit the following url\footnote{\label{foot1} {\tt http://utopia.duth.gr/~nmitiano/mdld.htm}} and listen to the described separation results.  The proposed MDLD approach seems to outperform our previous separation effort MoWL and Hyv\"arinen's algorithm in terms of all the performance indexes. The proposed MDLD approach is not susceptible to bordering effects, since it is circular by definition and avoids shortcomings of our previous offerings. Compared to a state-of-the-art method, such as ``GaussSep'', our method is better in terms of the SIR index but is falling behing in terms of the SDR and SAR indexes. The SIR index reflects the capability of an algorithm to remove interfence from other sources in the mixture. The SAR index refers to the audible artifacts that remain in the separated signals, due to the overlapping of several points in the time-frequency space (even in the MDCT representation) in the underdetermined mixture that are incorrectly attributed to either source. In this sense, our algorithm seems to perform slightly better compared to ``GaussSep'' in terms of removing ``crosstalk'' from other sources, but there seem to be more audible artifacts after separation in our approach compared to ``GaussSep''. This is due to the fact that the ``GaussSep'' segments the time-frequency representation in small localised neighbourhoods and performs local Gaussian Modelling so as to separate and filter sources from those areas that separation is more achievable. Instead, our approach simply clusters all time-frequency points according to the fitted DLD using hard thresholds (or soft-thresholds in the case $K>2$).
\begin{table}[htb]
\caption{Running time comparison with GaussSep and MoWL approaches. The measurements are in seconds.}
\begin{center}
{
\begin{tabular}[width=0.0.5\textwidth]{|l||c|c|c|}\hline
&\bf{MDLD}&\bf{Gaussep}&\bf{MoWL}\\\hline\hline
Groove&2.39&224.21&20.46\\\hline
Latino1&1.27&122.02&5.48\\\hline
Latino2&1.28&129.09&3.59\\\hline
Dev2Male3&2.31&72.64&19.67\\\hline
Dev2Female3&2.33&75.92&16.09\\\hline
Dev2WDrums&2.07&56.79&8.55\\\hline
Dev1WDrums&1.55	&54.06&11.88\\\hline\hline
\it{Average}&\bf{1.88}&\bf{104.96}&\bf{12.24}\\\hline\hline
Dev3Female3&9.56&1021.31&-\\\hline
Example(3$\times$5)&4.04&1598.7&-\\\hline
Example(4$\times$8)&9.393&2359.1&-\\\hline\hline
\it{Average}&\bf{7.66}&\bf{1659.70}&-\\\hline
\end{tabular}
}
\end{center}
\label{Table6}
\end{table}

Another important issue is to compare the processing time of the three best performing algorithms. All experiments were conducted on an Intel Core i5-460M (2.53 GHz) with 4GB DDR3 SDRAM running Windows Professional 64-bit and MATLAB R2011a. Our MATLAB implementations of the MDLD and MoWL algorithms were not optimised in terms of execution speed.  In Table \ref{Table6}, the typical running time in seconds is summarised for each experiment and method. The first observation is that the MDLD approach is  faster compared to our previous MoWL. As it was previously mentioned, employing a mixture of wrapped Laplacians to solve the ``circularity'' problem entails the running of two EM algorithms: one for the wrapped Laplacians and one for the mixture of wrapped Laplacians. This seems to delay the convergence of the algorithm. Instead, the MDLD requires the training of one EM algorithm for the mixture and even though is more complicated, it seems to converge faster compared to the MoWL. The second observation is that there is an important difference between the processing time of the MDLD approach and the ``GaussSep'' algorithm. As previously mentioned, the ``GaussSep'' algorithm is more complicated in structure thus justifying its long running time. Nevertheless, the proposed MDLD approach offers a very fast underdetermined source separation alternative with high SIR performance that can be used in environments where processing time is important. The third observation is that the processing time for the ``GaussSep'' algorithm scales significantly with the duration of the signals and the number of sources, i.e. the ``Groove'', ``Latino1'', ``Latino2'' (44.1KHz - 4 sources) require more time than the Dev2Male3 and Dev2Female3 sets (16KHz - 4 sources) and the Dev2WDrums and Dev1WDrums sets (16KHz - 3 sources). Instead, the MDLD's running time seems to be closer to the avarage in most cases, maybe slightly deteriorating with the complexity of the source separation problem.

\begin{table*}[htb]
\caption{The proposed MDLD approach is compared for source estimation performance ($K=3,4$) in terms of SDR (dB), SIR (dB) and SAR(dB) with the GaussSep approach. The measurements are averaged for all sources of each experiment.}
\begin{center}
{
\begin{tabular}[width=0.0.5\textwidth]{|l||c|c||c|c||c|c|}\hline
&\multicolumn{2}{c||}{\bf{SDR} (dB)}&\multicolumn{2}{c||}{\bf{SIR} (dB)}&\multicolumn{2}{c|}{\bf{SAR} (dB)}\\\hline
&MDLD&GaussSep&MDLD&GaussSep&MDLD&GaussSep\\\hline\hline
Dev3Female3&6.02&16.93&23.84&22.43&6.17&18.40\\\hline
Example $3\times5$ &3.91&9.94&17.92&15.21&4.17&11.68\\\hline
Example $4\times8$ &2.24&-18.63&16.4&-17.58&2.52&9.39\\\hline
\end{tabular}
}
\end{center}
\label{Table7}
\end{table*}

\subsubsection{Underdetermined source separation examples with more than two mixtures}

In this section, we employ the described generalised DLD approach to perform separation of $3\times L$ and $4\times L$ mixtures. The 2-mixtures setup, that dominates the literature, may also arise from the fact that most audio recordings and CD masters are available as stereo recordings (2 channels is equivalent to 2 mixtures), where we need to separate the instruments that are present. Nowadays, the music industry is moving towards multichannel formats, including the 5.1 and the 7.1 surround sound formats, which implies more than 2 channels will be available for processing. In this section, we will attempt to perform separation of the Dev3Female3 set from SiSEC2011~\cite{SiSEC2011} and a $3\times 5$ (3 mixtures - 5 sources) and a $4\times 8$ (4 mixtures - 8 sources) scenario using the male and female voices from Dev3. Our MDLD approach will be compared to the ``GaussSep'' algorithm that is able to work with multi-channel data. We used the same frame length  and time-frequency neighbourhood sizes for both algorithms as previously. The MDLD was initialised as described in the previous section. After fitting the model, we employed the soft-thresholding scheme, as it was described in~\cite{Mitianoudis07f}. Since it is not straightforward to calculate the intersection surfaces between the individual $p$-dimensional DLDs, we employ a soft-thresholding scheme, as described earlier. For our experiments, we used a value of $q=0.8$.

For the $3\times 5$ example, we centred the 5 speech sources around the angles $\theta_1= [0^o,-87^o,-60^o,0^o,45^o]$ and $\theta_2 = [85^o,0^o,-60^o,0^o,45^o]$.  The sources were mixed using the mixing matrix $\mathbf{A}=[\cos\theta_2\cos\theta_1 ; \cos\theta_2\sin\theta_1;\sin\theta_2]$. For the $4\times 8$ example, we centred eight audio sources around the angles: $\theta_1= [-75^o,-30^o,0^o,50^o,10^o,80^o, -45^o, 0^o]$, $\theta_2= [ 70^o,30^o,-20^o,50^o,-70^o,0^o,15^o,-70^o]$ and $\theta_3= [ 80^o,20^o,10^o,-50^o, 0^o,-10^o,-25^o, -35^o ]$. The sources were mixed using the mixing matrix $\mathbf{A}=[\cos\theta_3\cos\theta_2\cos\theta_1;\cos\theta_3\cos\theta_2\sin\theta_1;$ $\cos\theta_3\sin\theta_2;$ $\sin\theta_3]$.

 The separation results for the three experiments in terms of SDR, SIR and SAR can be summarised in Table \ref{Table7}. The reader can listen to the audio results from the following url (See Footnote \ref{foot1}). In the case of $K=3$ mixtures, both algorithms managed to perform separation in either case. Similarly to the $K=2$ case, the ``GaussSep'' featured higher SDR  and SAR performances, whereas the proposed MDLD featured higher SIR performance. The image is completely different in the case of $K=4$ mixtures, where the MDLD manages to separate all 8 sources in contrast to the ``GaussSep'' that fails to perform separation. This might be due to fact that the sparsest ML solution in the optimisation of \cite{Vincent09} is restricted to vectors with $K\leq 3$ entries, i.e. 3 sources present at each point. In contrast, the proposed MDLD algorithm is designed to operate for any arbitrary number of sensors $K$, without any constraint.

In Table \ref{Table6}, we can see the processing times for the two algorithms for the three experiments. The MDLD processing time has increased slightly but still remains relatively fast, requiring an average of 7.66 secs to perform separation. This implies that the computational complexity of the proposed MDLD algorithm does not scale considerably with the number of sources $L$ and sensors $K$. In contrast, the ``GaussSep'' algorithm's processing has increased considerably with $K$. The processing time seems to scale up dramatically with increasing $K$  and number of estimated sources $L$. For $K=3$, it required an average of 1310 sec and for $K=4$, it required 2359 sec which is almost the double processing time for $K=3$. Thus, it appears that the proposed MDLD algorithm is capable of offering a faster and more stable multichannel solution to the underdetermined source separation problem, featuring higher SIR rates, compared to a state-of-the-art approach.

The main aspiration for future work behind these experiments is to combine the speed and stability of the MDLD approach with the low-artifact separation quality, proposed by Vincent et al~\cite{Vincent09}. It might be possible to import this time-frequency localised source separation framework, where the source clusters can be modeled by mixtures of MDLDs. A more intelligent fuzzy clustering algorithm may combine the information from the MDLD priors to attribute points to multiple sources, overcoming the artifacts that arise from the partitioning of the time-frequency space.

\section{Conclusion}

In this paper, the problem of modelling multidimensional Directional Sparse data is addressed. This work is building on previous work on directional Gaussian models (i.e. the von-Mises and the vonMises-Fisher densities) to propose a novel generalised Directional Laplacian model for modelling multidimensional directional sparse data. Maximum Likelihood estimates of the densities' parameters were proposed along with an EM-algorithm that handles the training of DLD mixtures . The proposed algorithms were tested with randomly generated synthetic data where the algorithms demonstrated good performance in modelling the directionality of the data. The proposed algorithm can also offer a solution for the general multichannel   underdetermined source separation problem ($K\geq 2$), offering fast and stable performance and high SIR compared to state-of-the-art methods~\cite{Vincent09}.

For future work, the authors will look for methods to incorporate the time-frequency localised source separation framework~\cite{Arberet10,Vincent09}, in order to reduce the amount of audible artifacts in the separated sources. Another future direction is to adapt this technique for a convolutive-mixture scenario, where using the Short-Time Fourier Transform, we can transform the convolutive mixtures into multiple complex instantaneous mixtures. Source separation-clustering for each frequency bin can be performed using a modified version of the proposed algorithm and permutation alignment can be performed using Time-Frequency Envelopes or Direction-of-Arrival methods~\cite{Mitia02a,Mitia04e,Sawada04a}. The speed of the proposed MDLD algorithm can be very useful, since frequency-domain convolutive methods need to solve many complex instantaneous source separation problems simultaneously.

\section*{Acknowledgment}

The author would like to thank Dr. Laurent Daudet for providing the code for the MDCT analysis~\cite{Daudet04}. The author would like to thank the anonymous reviewers for their kind suggestions and corrections that helped to improve the quality of the paper.

\appendix
\subsection{Calculation of the normalisation parameter for the Generalised DLD}
\label{appxMD1}
To estimate the normalisation coefficient $c_p(k)$ of (\ref{MDDLD}), we need to solve the following equation:
\begin{equation}\nonumber
\int_{\mathbf{x}\in\mathcal{S}^{p-1}} c_p(k)e^{-k\sqrt{1-(\mathbf{m}^T\mathbf{x})^2}} d\mathbf{x}=1
\end{equation}
Following equation (B.8) and in a similar manner to the analysis in Appendix B.2 in~\cite{Dhillon03}, we can rewrite the above equation as follows:
\begin{equation}\nonumber
c_p(k)\int_{0}^{\pi}d\theta_{p-1}\int_{0}^{\pi} e^{-k\sqrt{1-\cos^2\theta_1}}\sin^{p-2}\theta_1 d\theta_1 \times
\end{equation}
\begin{equation}\nonumber
\times \prod_{j=3}^{p-1}\int_{0}^{\pi}\sin^{p-j}\theta_{j-1} d\theta_{j-1}=1
\end{equation}
Following a similar methodology to Appendix B.2 in~\cite{Dhillon03}, the above yields:
\begin{equation}\nonumber
c_p(k)\pi\int_{0}^{\pi} e^{-k\sin\theta_1}\sin^{p-2}\theta_1d\theta_1 \frac{\pi^{\frac{p-3}{2}}}{\Gamma(\frac{p-1}{2})}=1
\end{equation}
Using the definition of $I_p(k)$, we can write
\begin{equation}\nonumber
c_p(k)I_{p-2}(k)\frac{\pi^{\frac{p+1}{2}}}{\Gamma(\frac{p-1}{2})}=1\Rightarrow c_p(k)=\frac{\Gamma(\frac{p-1}{2})}{\pi^{\frac{p+1}{2}}I_{p-2}(k)}
\end{equation}
\subsection{Gradient updates for $\mathbf{m}$ and $k$ for the MDDLD}
\label{appxMD2}
The first order derivative of the log-likelihood in (\ref{LogLike_MDDLD}) for the estimation of $\mathbf{m}$ are calculated below:
\begin{eqnarray}
\frac{\partial J(\mathbf{X},\mathbf{m},k)}{\partial \mathbf{m}}&=&-k\sum_{n-1}^N\frac{-2\mathbf{m}^T\mathbf{x}_n}{2\sqrt{1-(\mathbf{m}^T\mathbf{x}_n)^2}}\mathbf{x}_n\nonumber\\
&=& k\sum_{n=1}^N\frac{\mathbf{m}^T\mathbf{x}_n}{\sqrt{1-(\mathbf{m}^T\mathbf{x}_n)^2}}\mathbf{x}_n
\end{eqnarray}

Before we estimate $k$ from the log-likelihood (\ref{LogLike_MDDLD}), we derive the following property:
\begin{equation}
\frac{\partial}{\partial k}I_0(k)= -\frac{1}{\pi}\int_{0}^{\pi} e^{-k\sin\theta}\sin\theta d\theta=-I_1(k)\nonumber
\end{equation}
The above property can be generalised as follows:
\begin{equation}
\frac{\partial^p}{\partial k^p}I_0(k)=(-1)^{p}\frac{1}{\pi}\int_{0}^{\pi}\sin^p\theta  e^{-k\sin\theta}d\theta=(-1)^{p}I_p(k)\nonumber
\end{equation}
The first order derivative of the log-likelihood in (\ref{LogLike_MDDLD}) for the estimation of $k$ are then calculated below:
\begin{equation}
\frac{\partial J(\mathbf{X},\mathbf{m},k)}{\partial k}=N\frac{I_{p-1}(k)}{I_{p-2}(k)}-\sum_{n=1}^{N}\sqrt{1-(\mathbf{m}^T\mathbf{x}_n)^2}
\end{equation}

\subsection{A Directional K-Means algorithm}
\label{appx4}
Assume that $K$ is the number of clusters, $\mathcal{C}_i,\textrm{ }i=1,\dots,K$ are the clusters, $\mathbf{m}_i$ are the cluster centres and  $\mathbf{X}=\{\mathbf{x}_1,\dots,\mathbf{x}_n,\dots,\mathbf{x}_N\}$  is a $p$-dimensional angular  dataset lying on the half-unit $p$-D sphere. The original K-means~\cite{MacQueen67} minimises the following non-directional error function:
\begin{equation}
Q=\sum_{n=1}^N\sum_{i=1}^K||\mathbf{x}_n-\mathbf{m}_i||^2
\end{equation}
where $||\cdot||$ represents the Euclidean distance. Instead of using the square Euclidean distance for the $p$-dimensional Directional K-Means, we introduce the following distance function:
\begin{equation}
D_l(\mathbf{x}_n,\mathbf{m}_i)=\sqrt{1-(\mathbf{m}_i^T\mathbf{x}_n)^2}
\end{equation}
The novel function $D_l$ is similarly monotonic as the original distance but emphasizes more the contribution of points closer to the cluster centre. In addition, $D_l$ is periodic with period $\pi$. The $p$-dimensional Directional K-Means can thus be described as follows:
\begin{enumerate}
\item
Randomly initialise $K$ cluster centres $\mathbf{m}_i$, where $||\mathbf{m}_i||=1$
\item
Calculate the distance of all points  $\mathbf{x}_n$ to the cluster centres $\mathbf{m}_i$, using $D_l$.
\item
The points with minimum distance to the centres $\mathbf{m}_i$ form the new clusters $\mathcal{C}_i$.
\item
The clusters $\mathcal{C}_i$ vote for their new centres $\mathbf{m}_i^+$. To avoid averaging mistakes with directional data, vector averaging is employed to ensure the validity of the addition. The resulting average is normalised to the half-unit $p$-dimensional sphere:
\begin{equation}
\mathbf{m}_i^+=\frac{1}{C_i}\sum_{\mathbf{x}_n \in C_i }\mathbf{x}_n
\end{equation}
\begin{equation}
\mathbf{m}_i^+\leftarrow\mathbf{m}_i^+/||\mathbf{m}_i^+||
\end{equation}
\item
Repeat steps 2), 3), 4) until the means $\mathbf{m}_i$ have converged.
\end{enumerate}

\ifCLASSOPTIONcaptionsoff
  \newpage
\fi

%

%

\bibliographystyle{IEEEbib}
\bibliography{ica}

\end{document}